\documentclass{ws-procs9x6}

\begin{document}

\title{Hard Rescattering Mechanism in High Energy Photodisintegration of 
the Light Nuclei}
\author{M.M.~Sargsian}

\address{Department of Physics,  Florida International University 
Miami, FL 33199 }  


\maketitle

\abstracts{
We discuss the high energy photodisintegrataion of light nuclei 
in which the energy of the absorbed photon is equally shared between 
two nucleons in the target. For these reactions we investigate the 
model in which photon absorption by a quark in one nucleon followed by 
its high momentum transfer interaction with a quark of the other nucleon
leads to the production of  
two nucleons with high relative momentum. We sum the relevant quark 
rescattering diagrams, and demonstrate that the scattering amplitude 
can be expressed as a convolution of the large angle $NN$ scattering amplitude, 
the hard photon-quark interaction vertex and the low-momentum 
nuclear wave function. Within this model we calculate the 
cross sections and polarization 
observables  of high energy $\gamma +d\rightarrow pn$ and 
$\gamma + ^3He\rightarrow pp +n$ reactions.}

\section{Introduction}
The reactions of  high energy photodisintegration of light nuclei 
in which the energy of the incoming photon is equally shared by two 
outgoing nucleons is unique in a sense that it deposits to 
the nuclear system a large amount of energy which should be 
shared at least by two nucleons in the nuclei (see e.g.\cite{BCh,Holt}). 
When this  energy  exceeds the typical meson mass ({\em sevearl hundred MeV}) 
one expects that the interaction picture based on a meson exchange currents should 
break down. In this limit the hard contribution is expected to dominate.
As a result one expects that the quark gluon picture of interaction will become a 
relevant framework for description of the reaction.

The experiments on high energy two-body photodisintegration of the
deuteron\cite{E89012,NE17} demonstrated that starting at  
E$_\gamma \ge$ 1 GeV the conventional mesonic picture 
of nuclear interactions is indeed breaking down.  One of the first predictions for 
$\gamma d \rightarrow pn$ reactions  within QCD was based on  the quark counting rule, 
which predicted ${d\sigma/dt\sim s^{-11}}$.
This prediction was based on the hypothesis that the Fock state 
with the minimal number of partonic constituents will dominate in 
two-body large angle hard collisions\cite{hex}. Although successful in 
describing energy dependences of number of hard processes, this
hypothesis does not allow to make calculation of the absolute 
values of the cross sections. 
Especially for reactions involving baryons, the calculations within 
perturbative QCD underestimate  the measured cross sections by orders 
of magnitude see e.g.\cite{Isgur_Smith}. 
This may be an indication that in the accessible range of energies 
bulk of the interaction is in the domain of the nonperturbative
QCD\cite{Isgur_Smith,Rady}, for which the theoretical methods of 
calculations are very limited.

\vspace{-0.4cm}
\begin{figure}[htb]
\centerline{\epsfxsize=3.5in\epsfbox{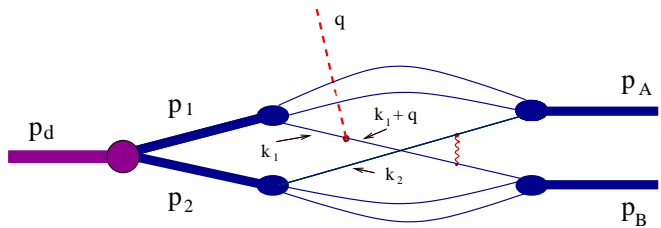}} 
\caption{Quark Rescattering diagram.}
\label{Fig.1}
\end{figure}
\vspace{-0.4cm}

\section{Hard Rescattering Model} 
Recently we suggested a model in which the absorption of the photon
by a quark of one  nucleon, followed by  a high-momentum transfer
(hard) rescattering with a quark from the second nucleon, produces the 
final two nucleon state of large relative momenta. The typical
diagram representing such a scenario is presented in Figure 1.
Based on the analysis of these type of diagrams we find that:
\begin{itemize}
\item the dominant contribution comes from the soft vertices of $d\rightarrow NN$ transition,
while quark rescattering  proceeds trough hard gluon exchange
\item the  $d\rightarrow NN$ transition can be evaluated  based on the 
conventional deuteron wave function calculated using the realistic nucleon-nucleon potentials. 
\item the structure of hard interaction for the rescattering part of 
the reaction is similar to that of hard NN scattering.
\item as a result the sum of the multitude of diagrams with incalculable 
nonperturbative parts of the interaction can be  expressed through 
the experimentally measured amplitude of hard $NN$ scattering.
\end{itemize}

\section{Kinematics and the  Cross Section}

We consider the kinematics of sufficiently large energies
$E_{\gamma}\ge 2 ~GeV$ and momentum transfer
$-t,-u\ge 2~GeV^2$. These restrictions allow to make a two important 
approximations: first is that the mass of the  intermediate hadronic state
produced in the $\gamma N$ interaction is in deep inelastic continuum, 
thus the partonic picture is relevant. The second one is that the quark 
rescattering is hard and can be factorized from the soft nuclear wave function. 
We evaluate Feynman diagrams such as Fig.~1, in which quarks are exchanged 
between nucleons via the exchange of a gluon. All other quark-interactions
are included in the  partonic wave function of the nucleon, $\psi_N$.
We use  a simplified notation in which only the momenta of the
interacting quarks require labeling.
The scattering amplitude $T$ for photo-disintegration of the  deuteron
(of four-momentum $p_d$ and  mass $M_d$) into two nucleons of momentum
$p_A$ and $p_B$ is given by:  
\begin{eqnarray}
&& T = -\sum\limits_{e_q} \int  \left( 
{\psi_N^\dag(x'_2,p_{B\perp},k_{2\perp})\over x'_2}\bar u(p_B-p_2+k_2)\right. 
\left[-igT_c^{F}\gamma^{\nu}\right]
\nonumber \\
&& {u(k_1+q)\bar u(k_1+q)\over (k_1+q)^2 - m_q^2 + i\epsilon}
\left[-ie_q\epsilon^{\perp}\cdot\gamma^{\perp}\right]
\left. u(k_1)
{\psi_N(x_1,p_{1\perp},k_{1\perp})\over x_1}\right)
\nonumber \\
&& \left\{{\psi_N^\dag(x'_1,p_{A\perp},k_{1\perp})\over x'_1}
\bar u(p_A-p_1+k_1)\left[-igT_c^{F}\gamma_{\mu}\right]
u(k_2) {\psi_N(x_2,p_{2\perp},k_{2\perp})\over x_2}\right\}
\nonumber \\
&& G^{\mu\nu} {\Psi_{d}(\alpha,p_{\perp})\over 1-\alpha}
{dx_1\over 1-x_1} {d^2k_{1\perp}\over 2(2\pi)^3}
{dx_2\over 1-x_2} {d^2k_{2\perp}\over 2(2\pi)^3}
{d\alpha\over \alpha} {d^2p_\perp\over 2(2\pi)^3},
\label{Ta}
\end{eqnarray}
where $p_1$ and  $p_2$ are the momenta of the nucleons in the deuteron,
with $\alpha \equiv {p_{1+}\over p_{d+}}$, $p_2=p_d-p_1$ and $p_{1\perp}=
-p_{2\perp}\equiv p_\perp$.
Each nucleon consists of one active  quark of momenta $k_1$  and $k_2$: 
$x_i\equiv {k_{i+}\over p_{i+}} = {k_{i+}\over \alpha p_{d+}}$ ($i=1,2$).
$G^{\mu\nu}$ describes the gluon exchange between interchanged quarks.
We use the reference frame where $p_d = (p_{d0},p_{dz},p_{\perp})\equiv
({\sqrt{s'}\over 2}+{M_d^2\over 2\sqrt{s'}},
{\sqrt{s'}\over 2}-{M_d^2\over 2\sqrt{s'}} ,0)$,
with  $s = (q+p_d)^2$, $s'\equiv s-M_D^2,$ and the photon four-momentum is
$q = ({\sqrt{s'}\over 2}, -{\sqrt{s'}\over 2},0)$.

We first observe that the denominator of the knocked-out 
quark propagator, when recoil quark-gluon system with mass $m_R$ is on 
mass shell, has a pole at 
$\alpha_c \equiv {x_1 m_R^2+k_{1\perp}^2\over (1-x_1)x_1\tilde s}$: 
\begin{equation} 
(k_1+q)^2-m_q^2 + i\epsilon \approx x_1s^\prime(\alpha-\alpha_c + i\epsilon),
\label{alphac}
\end{equation}
where  $\tilde s \equiv s'(1+{M_d^2\over s'})$.
Next we calculate the photon-quark hard scattering vertex 
and integrate over the $\alpha$ using 
only the pole contribution in Eq.(\ref{alphac}). 
Note that the dominant 
contribution arises from the soft component of the deuteron when
$\alpha_c={1\over 2}$, which requires 
$k_{1\perp}^2\approx {(1-x_1)x_1\tilde s\over 2}$.

Summing over the struck quark contributions from
photon scattering off neutron and proton  one can express
the scattering amplitudes through the $pn$ hard scattering amplitude  
within the quark interchange mechanism~(QIM)-$A_{pn}^{QIM}(s,l^2)$ as follows:
\begin{eqnarray}
T \approx   
{ie (\epsilon^++\epsilon^{-})(e_u + e_d) \over  2 \sqrt{s'}}
\int f({l^2\over s})A_{pn}^{QIM}(s,l^2) \Psi_{d} ({1\over 2},p_{\perp})
{d^2p_\perp\over (2\pi)^2}.
\label{Tc}
\end{eqnarray}
where  $l = p_p - p_1$, 
$\epsilon^\pm={1\over \sqrt{2}}(\epsilon_x\pm i \epsilon_y)$ and 
$e_u$ and $e_d$ are the electric charges of $u$ and $d$ quarks.
The factor $f(l^2/s)$ accounts for the difference between  the hard
propagators  in our process and those occurring in wide angle $pn$
scattering. Within the  Feynman mechanism\cite{Feynman},  the interacting 
quark carries the whole momentum of the nucleon  ($x_{1}\rightarrow 1$), 
thus $f(l^2/s)=1$.
Within the minimal Fock state approximation, $f(l^2/s)$ 
is the scaling function of the $\theta_{cm}$ only with 
$f(\theta_{cm}=90^0)\approx 1$\cite{gdpn}.

We compute the differential cross section averaging $|T|^2$ over the spins
of initial photon and deuteron and summing over the spins of the  final
nucleons. Then we use the observation that the quark interchange topologies
are the dominant for fixed angle, $\theta_{cm}=90^0$ high
momentum transfer (non strange) baryon-baryon 
scattering. Thus in the region of $\theta_{cm}\approx 90^0$ we
replace $A_{pn}^{QIN}$ by the experimental data - $A_{pn}^{Exp}$ and 
obtain\cite{gdpn}\footnote{The Eq.(6) in Ref.\cite{gdpn} contains a 
misprint: $4$ in front of Eq.(6) should be $8$. Note that 
the numerical calculations in \cite{gdpn} are done with the correct factor.}:
\begin{eqnarray}
{d\sigma^{\gamma d\rightarrow pn }\over dt}  =  {8\alpha\over 9}\pi^4\cdot
{1\over s'} C({\tilde t\over s}){d\sigma^{pn\rightarrow pn}(s,\tilde t)\over dt}
\nonumber \\
\times \left| \int\Psi_d^{NR}(p_z=0,p_\perp)\sqrt{m_N}
{d^2p_\perp\over (2\pi)^2}\right|^2,
\label{difcrsb}
\end{eqnarray}
where $\tilde t = (p_B-p_d/2)^2$. Here $\Psi_d^{NR}$ is the nonrelativistic 
deuteron wave function which can be calculated using realistic NN interaction 
potentials. The function $C({\tilde t\over s})\approx f^2(\tilde t/s)\approx 1$
at $\theta_{cm}\sim 90^0$ and  is slowly varying function of $\theta_{cm}$.

Within the hard rescattering model we can calculate also the high energy 
photodisintegration of $^3He$ in the specially chosen kinematics, 
in which the absorbed photon's energy is equally shared between the two outgoing 
protons while the  final neutron is very slow $p_n\le 100 MeV/c$. 
In this case the cross section of the reaction is 
expressed through the nonrelativistic wave function of the  $^3He$ and 
the amplitude of hard $pp$ scattering, with small corrections coming from 
soft $pn$ rescattering.

\begin{figure}[htb]
\centerline{\epsfxsize=3.6in\epsfbox{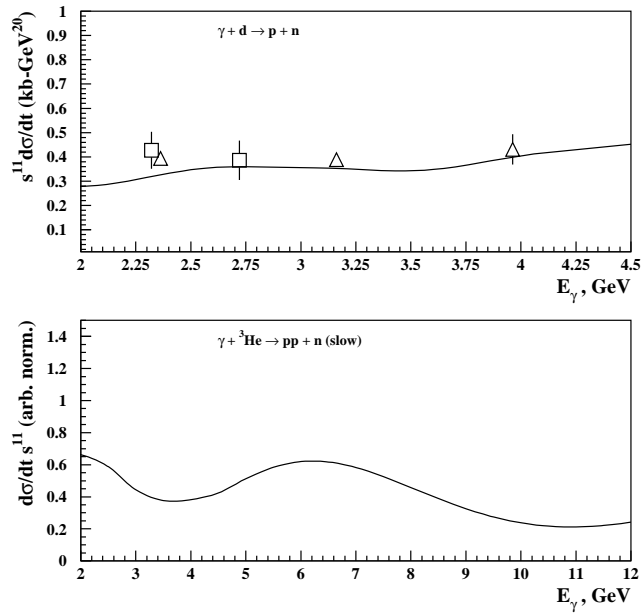}} 
\caption{The ${d\sigma/dt}s^{11}$ as a function of $E_{\gamma}$ at $\theta_{cm}=90^0$.  
Data are from [3] (triangles) and [4] (squares).}
\label{Fig.2}
\end{figure}

\section{Comparison with the Data}
\subsection{Energy dependence}

One can estimate the energy dependences of the differential cross sections 
of the $\gamma d\rightarrow pn$ and $\gamma ^3He\rightarrow pp +n$  reactions
at $\theta_{cm}=90^0$ using the experimental data on hard $pn$ and $pp$ 
scattering. Figure 2 represents the calculations and the comparison 
with existing data on deuteron target. Our calculation of 
$\gamma d\rightarrow pn$ has $\approx 25\%$ accuracy because of the 
rather large experimental errors in high momentum transfer $pn$ scattering
cross section.  The prediction for $\gamma ^3He\rightarrow pp +n$ cross 
section (which should be considered preliminary) demonstrates the
oscillation which is related to the observed oscillations in high 
momentum transfer $pp$ cross section\cite{RP}.

\subsection{Angular Dependence}

As we mentioned above, the hard rescattering mechanism can predict the 
value of $C(t/s)\approx 1$ only at $\theta_{cm}=90^2$. One expects that 
this function should be a smooth function of  $\theta_{cm}$. As a result 
the angular dependence of photodisintegration reaction should 
in general reflect the angular dependence of hard $NN$ scattering cross 
section. 

In particular it is interesting to observe that the $pn$ scattering cross 
section exhibits a strong  angular asymmetry with the cross sections 
being dominate at forward angles.  As Figure 3 demonstrates, within 
hard rescattering model this feature is  reflected also in the angular 
dependence of  $\gamma d\rightarrow pn$ cross section. This result is  
in a qualitative agreement with the preliminary JLab data.
 
\vspace{-0.4cm}
\begin{figure}[htb]
\centerline{\epsfxsize=3.5in\epsfbox{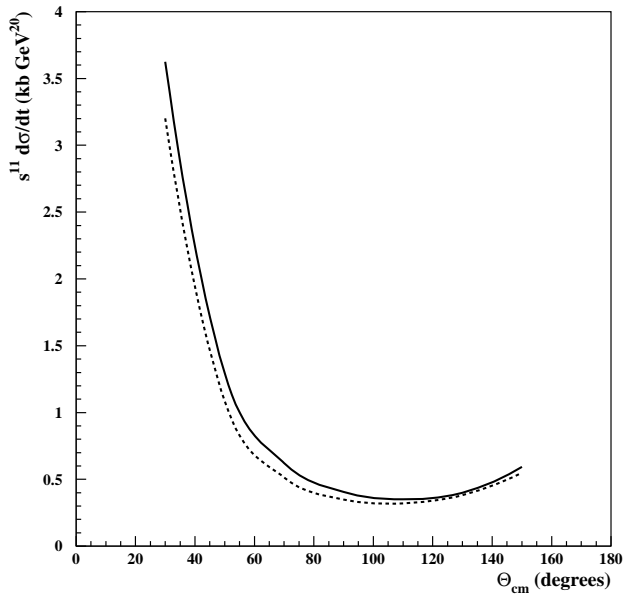}} 
\caption{Angular dependence, for $\gamma d\rightarrow pn$ reaction at $E_{\gamma}=2 GeV$ 
(dashed curve) and $3~GeV$ (solid curve).}
\label{Fig.3}
\end{figure}


\section{Polarization Observables}
Hard rescattering model allows also to calculate the different 
polarization observables of the photodisintegration reaction.
Writing down explicitly the helicity indexes of interacting 
particles, for the scattering amplitude of Eq.(\ref{Tc}) one obtains:
\begin{eqnarray}
& &  T^{h,m,\eta_2,\eta_1}  \approx   \\ \nonumber
& & \ \ \ \ {i (e_u+e_d)\over  2 \sqrt{s'}}
\int f({l^2\over s})<\eta_1,\eta_2|A_{pn}(s,l^2)|h,\lambda_2> 
\Psi^{m,h,\lambda2}_{d} ({1\over 2},p_{\perp}){d^2p_\perp\over (2\pi)^2}.
\label{Tcpol}
\end{eqnarray}
where $h$, $m$, $\eta_2$ and $\eta_1$ represent the helicities of 
incoming photon, target deuteron and two outgoing nucleons respectively. 
In derivation of Eq.(\ref{Tcpol}) it is assumed additionally that the 
quark which interacts with the photon carries the total helicity of the 
parent nucleon. The  helicity amplitudes, $A_{pn}$  in Eq.(\ref{Tcpol}) can 
be expressed through the five independent helicity amplitudes:
\begin{eqnarray}
\phi_1(s,t) = <++|A|++> \ \ \ \ \phi_2(s,t) = <++|A|--> \nonumber \\
\phi_3(s,t) = <+-|A|+->  \ \ \ \ \phi_4(s,t) = <+-|A|-+> \nonumber \\
\phi_5(s,t) = <++|A|+->
\label{phis}
\end{eqnarray}
for which the following hierarchy relation can be stated in a model independent 
way\cite{RS}:
\begin{equation}
\phi_2 < \phi_5 < \phi_1,\phi_3,\phi_4
\label{rel}
\end{equation}
Using the definitions of Eq.(\ref{phis}) one can calculate the different 
asymmetries in the deuteron photodisintegration reaction. We are particularly 
interested in the recoil transverse $P_y$  and longitudinal $C_x$, $C_y$  
polarizations for which the high energy data are becoming available from 
Jefferson Lab\cite{Wij}. Based on Eq.(\ref{Tcpol}) the predictions for these 
observables are as follows:
\begin{eqnarray}
& & P_y = - 2{ I}m
\left\{\left[2(\phi_1(s,\tilde t)+\phi_2(s,\tilde t)) +\phi_3(s,\tilde t) - 
\phi_4(s,\tilde t)\right]\phi_5^{\dag}(s,\tilde t)\right\}/F(s,\tilde t) \nonumber \\ 
& & C_x = + 2{ R}e
\left\{\left[2(\phi_1(s,\tilde t)-\phi_2(s,\tilde t)) +\phi_3(s,\tilde t) + 
\phi_4(s,\tilde t)\right]\phi_5^{\dag}(s,\tilde t)\right\}/F(s,\tilde t) \nonumber \\ 
& & C_z = 0
\label{pols}
\end{eqnarray}
where 
$F(s,\tilde t) = 2|\phi_1(s,\tilde t)|^2 +  2|\phi_2(s,\tilde t)|^2
+ |\phi_3(s,\tilde t)|^2 +  4|\phi_4(s,\tilde t)|^2+
6|\phi_5(s,\tilde t)|^2$.
In these derivations we neglected by the $D$ wave contribution of the deuteron wave 
function, which is justified on the basis that the nucleon momenta which enter in the 
deuteron wave function in the integral of Eq.(\ref{Tcpol}) is restricted, $(\le 300 MeV/c)$.

Based on the relation of Eq.(\ref{rel}) and the fact that  for the  on-shell amplitude 
$\phi_5=0$ at $\theta_{cm}=90^0$ one can conclude in a rather model independent way that the  
$P_y$ and $C_x$ should be  small at $\theta_{cm}=90^0$ at $E_{\gamma}\ge 2 GeV$.
This result seems  in qualitative agreement with the available data and the planned 
experiment at Jefferson Lab in $E_{\gamma}\ge 2 GeV$ region  will allow check these 
predictions in more detail. 

It is interesting to note that the smallness of the 
polarization observables  predicted in the model is not related to the commonly used 
assumption of the applicability of pQCD and related only to the fact that the considered
asymmetries are determined by the small helicity component of $NN$ amplitude, $\phi_5$.

\medskip

\section{Summary and Outlook}

The underlying hypothesis of hard rescattering model is that the dynamics  
of the photoproduction reaction is determined by the physics of high-momentum 
transfer contained in the hard scattering NN amplitude. As a result the 
short-distance aspect of the deuteron wave function is not important, which allows 
us to use the conventional deuteron wave function in numerical calculations.  
This  hypothesis, if confirmed by additional studies, 
may suggest the existence of new type of calculable hard nuclear reactions in which 
the sum of the ``infinite'' number of quark interactions could be replaced by the
hard amplitude of $NN$ interaction.

The prediction of the model agrees reasonably well with the existing data on 
high energy photodisintegration of the deuteron at $\theta_{cm}=90^0$. 
More data,  especially with a two energetic proton final state 
in  $\gamma+^3He\rightarrow pp$ (high $\ p_t$) + n $(p_t\approx 0$) reaction and a
more detailed angular distribution would definitely allow  to verify this hypothesis.
The polarization measurement also will be crucial. It is very important to extend them 
in the energy region where anomalies are  observed in the polarized  hard $pp$ scattering.

\end{document}